\documentclass[floatfix,prd,lengthcheck,showpacs,amssymb,amsmath,amsfonts,aps,altaffilletter,nofootinbib,nopreprintnumbers,showpacsm]{revtex4-1}
 \usepackage[utf8]{inputenc}
\usepackage{color}
\usepackage{graphicx}
\usepackage{float}
\usepackage{subfigure}
\usepackage{amsmath}
\usepackage{acronym}
\usepackage{multirow}
\usepackage{tabu}
\usepackage{tikz}
\usetikzlibrary{arrows,shapes,trees,decorations.pathreplacing}
\usepackage{import}
\usepackage{svn-multi}
\usepackage{hyperref}
\hypersetup{
    colorlinks=true,
    linkcolor=blue,
    filecolor=magenta,      
    urlcolor=cyan,
}
\begin{document}

\title[]{PyCBC Live: Rapid Detection of Gravitational Waves from Compact Binary Mergers}
\author{Alexander H. Nitz$^1$, 
        Tito Dal Canton$^2$,
        Derek Davis$^3$,
        Steven Reyes$^3$
%        Duncan Brown$^3$,
        }

\address{$^1$ Max-Planck-Institut f{\"u}r Gravitationsphysik,
         Albert-Einstein-Institut, D-30167 Hannover, Germany}
\address{$^2$ NASA Postdoctoral Program Fellow, Goddard Space Flight Center, Greenbelt, MD 20771, USA}
\address{$^3$ Syracuse University, Syracuse, NY 13244, USA}
\date{\today}

\begin{abstract}
We introduce an efficient and straightforward technique for rapidly detecting gravitational waves from compact binary mergers. We show that this method achieves the low latencies required to alert electromagnetic partners of candidate binary mergers, aids in data monitoring, and makes use of multidetector networks for sky localization. This approach was instrumental to the analysis of gravitational-wave candidates during the second observing run of Advanced LIGO, including the period of coincident operation with Advanced Virgo, and in particular the analysis of the first observed binary neutron star merger GW170817, where it led to the first tightly localized sky map ($31~\mathrm{deg}^2$) used to identify  AT 2017gfo. Operation of this analysis also enabled the initial discovery of GW170104 and GW170608 despite non-nominal observing of the instrument.
\end{abstract}

\maketitle

\section{Introduction}
\label{s:intro}

The era of gravitational-wave multi-messenger astronomy began with the near simultaneous detection of a binary neutron star merger through gravitational waves (GW170817)~\cite{TheLIGOScientific:2017qsa} and gamma rays (GRB 170817A)~\cite{Goldstein:2017mmi}, spawning numerous astronomers and telescopes to perform observations~\cite{GBM:2017lvd}, and led to the first identification of an optical counterpart to gravitational-waves (AT 2017gfo)~\cite{Coulter:2017wya}. This single observation has confirmed long held suggestions that neutron star mergers can produce short gamma-ray bursts~\cite{1986ApJ...308L..43P,1986ApJ...308L..47G,elp1989,npp1992} and r-process fueled kilonova~\cite{lp1998,ros2005,mmd+2010}.

In this paper we present a new method to detect gravitational-wave events, employed during the second observing run of Advanced LIGO (O2) and during the coincident observation with Virgo in 2017, known as \texttt{PyCBC Live}. This analysis is designed to find gravitational waves from binary neutron star, neutron star-black hole, and binary black hole mergers in Advanced LIGO and Virgo data for the purpose of alerting astronomers to perform observational follow-up. Existing methodologies to search for gravitational-waves in low latency rely on elaborate use of multi-rate filtering~\cite{Adams:2015ulm,Messick:2016aqy,Marion:2004,Buskulic:2010zz,Cannon2011Early,Cannon2010}, basis construction by means of singular value decomposition~\cite{Messick:2016aqy}, or even decomposition into a series of infinite impulse response filters~\cite{Hooper:2011rb}. We show that reuse of the existing infrastructure for doing straightforward frequency-domain matched filtering used by the full deep ~\texttt{PyCBC}-based offline analysis~\cite{Usman:2015kfa,Nitz:2017svb,Canton:2014ena} is similarly able to achieve high-performance, low-latency identification of gravitational-waves, with a substantially simpler architecture and approach.

At the time of this writing, in addition to the binary neutron star merger GW170817, three binary black hole mergers have been reported by Advanced LIGO and Virgo~\cite{GW170814,GW170104,Abbott:2017gyy} during the O2 observing run. Notably, GW170104 and GW170608 were identified using \texttt{PyCBC Live} during adverse observational conditions~\cite{GW170104GCN,GW170608GCN}. GW170814 was detected during standard autonomous analysis~\cite{GW170814GCN}. We note that the final significance statements for events are typically made using deep offline analysis such as those discussed in \cite{Messick:2016aqy,Usman:2015kfa,Nitz:2017svb}. Unlike low latency analysis, these are able to employ more comprehensive data quality information~\cite{Nuttall:2018xhi,TheLIGOScientific:2017lwt,Massinger:2016yvc} and updated instrumental calibration~\cite{Cahillane:2017vkb}. The initial identification of GW170817 was based on detection in the LIGO Hanford observatory alone using the method presented in~\cite{Messick:2016aqy, GCN21505}. As a single detector analysis alone cannot provide a well constrained sky localization, the precise localization, which enabled the identification of the optical counterpart to GW170817, required the rapid analysis of data from the LIGO and Virgo observatories by \texttt{PyCBC Live}~\cite{GCN21513} and subsequent localization using the \texttt{BAYESTAR} algorithm~\cite{Singer:2015ema}.

In this paper, we discuss how \texttt{PyCBC Live} extracts gravitational-wave candidates under the requirement to produce results in low latency, how these candidate events are ranked, and how their statistical significance is estimated. We show that this technique can produce alerts with accurate sky localizations using the information from multiple detectors, which was demonstrated during the analysis of GW170817. Furthermore, using data from the first observing run of Advanced LIGO as a test bed, we demonstrate that the sensitivity of this analysis compares favorably to the full deep~\texttt{PyCBC}-based offline analysis.  We will discuss applications of this analysis to searching for gravitational waves from a single detector along with future enhancements. We note that during the O2 observing period, this analysis operated primarily on data from LIGO Hanford and LIGO Livingston when extracting signals and determining their significance. We will discuss the analysis in this context, however, the procedures described here are generalizable to additional observatories, which will be the configuration used in future observing runs.

\begin{figure*}[t]
  \centering
    \includegraphics[width=\textwidth]{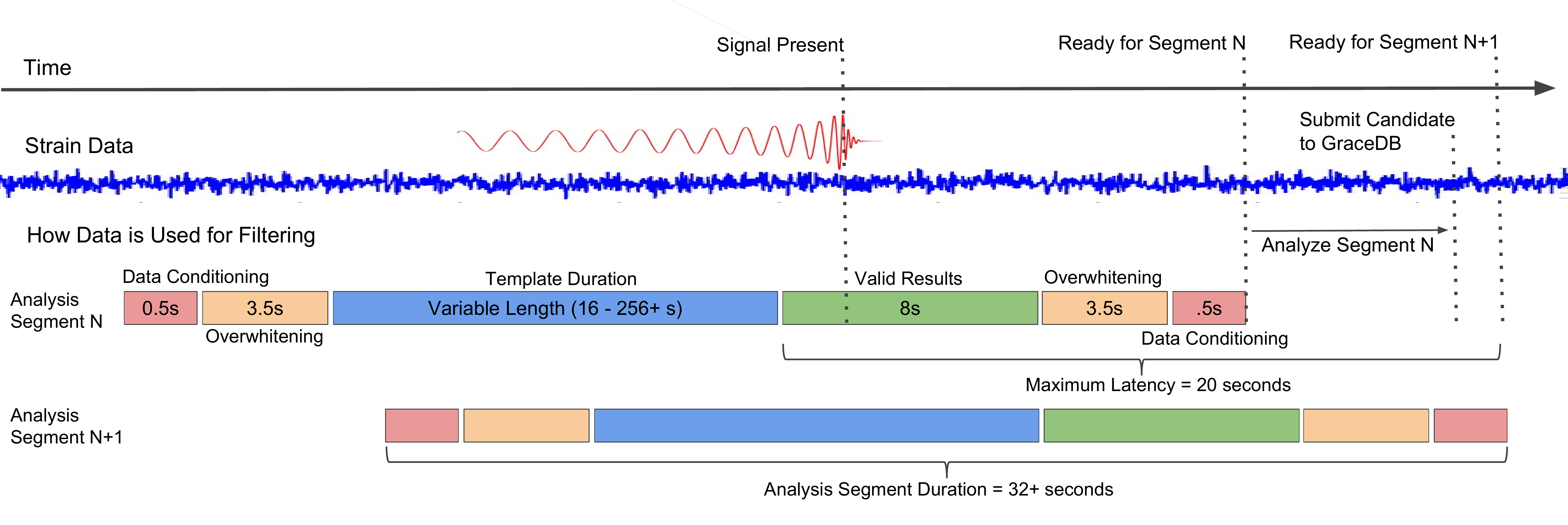}
\caption{A diagram of how data is processed by the \texttt{PyCBC Live} low-latency 
analysis. Data is analyzed in discrete \textit{analysis segments} with a fixed \textit{analysis stride} between them, in this case 8 seconds. The amount of data required varies by the template being analyzed. The maximum latency of this method is 20s for the configuration used in the diagram above. The average latency from merger to identification is 16s. In practice, additional time is required to calibrate and distribute the strain data to computing resources where the low latency analysis takes place, typically$~\sim$10s. Furthermore, critical information such as the sky localization, nearly takes an additional 10s to be generated after an event is initially identified and uploaded to the Gravitational-wave Candidate Database (GraceDB)~\cite{GraceDB}. This occurs as a separate follow-up process.}
\label{fig:flow}
\end{figure*}

\section{Methodology}

The most sensitive searches for gravitational waves from compact binary mergers of LIGO/Virgo data have relied upon matched filtering~\cite{Usman:2015kfa,Wiener:1949,Cutler:1992tc,Allen:2005fk}, which is an optimal technique for extracting signals from stationary colored Gaussian noise. Since the data is not Gaussian, but instead contains many classes of transient non-Gaussian noise, numerous additional signal consistency tests are also employed~\cite{Allen:2005fk} to mitigate the large signal-to-noise ratios that can be produced~\cite{TheLIGOScientific:2016zmo,Nuttall:2015dqa}. This method works when there are accurate models of the gravitational waveform~\cite{Blanchet:2002av,Faye:2012we,Taracchini:2013rva,Bohe:2016gbl}. A discrete set of waveform templates is chosen carefully to search over a wide range of binary component masses and spins~\cite{Sathyaprakash:1991mt,Babak:2006ty,Ajith:2012mn,Brown:2012qf}. The parameter space searched and the templates chosen for the first and second observing runs of Advanced LIGO are described in~\cite{TheLIGOScientific:2016qqj,DalCanton:2017ala}.

In this section we will detail how we adapt these methods for low latency analysis to extract candidate events from gravitational-wave data (\ref{sec:extract}), rank candidates found in one or more detectors (\ref{sec:rank}), and empirically estimate the background to determine the rate of false alarms (\ref{sec:bkg}). We find that we are able to achieve a high throughput low-latency analysis, using only straightforward frequency-domain matched-filtering, based on the methods described in \cite{Usman:2015kfa} and \cite{Allen:2005fk}. The philosophy behind the design of this analysis has been to achieve low latencies while maintaining a straightforward filtering design and leveraging the tools built into the PyCBC gravitational-wave analysis library~\cite{pycbc-github} which was used to construct the deep offline analysis that determines the significance of LIGO/Virgo detections~\cite{Usman:2015kfa,Nitz:2017svb}.

The latency goal of \texttt{PyCBC Live} is determined by external latencies such as the production, aggregation and transfer of the strain data, the sky localization of candidates, and the generation and distribution of the resulting alerts. All these processes set the appropriate latency time scale to achieve at O(10s), which is achieved by \texttt{PyCBC Live} as described next.

\subsection{Extracting Gravitational-wave Signals}
\label{sec:extract}
% Discuss a bit on the low latency dq information and what / how it is used

This section focuses on the extraction of gravitational-wave signals from observatory data: the aim of this step is to generate a matched-filter signal-to-noise-ratio (SNR) time series for each template in a predefined bank of waveforms. Each template represents a specific choice of signal parameters the analysis is searching for. Low-latency design requires that the analysis of a portion of data be completed for the entire template bank within a short, fixed amount of time, and be able to keep pace with the incoming data. This portion of the analysis is highly parallel and may conceptually proceed independently for each template and each observatory's data being analyzed. Finite impulse response (FIR) filters are used to guarantee a fixed latency with predetermined time invariant properties. As such, the overall latency of our analysis is limited by the duration of all the filtering operations, along with the \textit{analysis stride}, which determines the pace at which data is analyzed. For the filter configuration used in the O2 LIGO/Virgo observing run, the maximum latency incurred by this analysis method is 20s, with an average latency of only 16s.

In O2 the~\textit{analysis stride} was chosen to be 8 seconds. New data is read in and preconditioned by highpassing and resampling using zero-phase FIR filters generated by applying a Kaiser window to the ideal frequency response~\cite{scipy}. The highpass filter corner frequency is set at 15 Hz, primarily to reduce the dynamic range of the data. This allows the computationally-intensive part of the analysis to be performed in single-precision floating point arithmetic. We also downsample the data, which is originally recorded at 16384 Hz, to 2048 Hz. We can safely discard the higher frequency data as the bulk of the signal power is accumulated well below the Nyquist frequency of 1024 Hz, even for binary neutron star signals which reach several kHz. The portion of time-domain data corrupted due to filtering is then discarded and a continuous stream of preconditioned data, free from boundary artifacts, is produced by combining with previous data. The preconditioned stream is delayed by $\sim0.5$ s due to the chosen duration of the highpass and resampling FIR filters.

The matched filter signal-to-noise ratio $\rho$ is constructed according to~\cite{Cutler:1992tc,Allen:2005fk}, for a template waveform $h$, and data $s$, which can be defined as,
\begin{equation}
 \rho^2 \equiv \frac{\|\left\langle s | h \right\rangle\|^2}{\left\langle h | h \right\rangle}
\end{equation}
\label{eq:snr}
where the inner product is
\begin{equation}
 \langle a|b\rangle = 4 \int^\infty_0 \frac{\tilde{a}(f)\tilde{b}^*(f)}{S_n(f)} df
\end{equation}
and $S_n(f)$ is the estimated one-sided power spectral density (PSD) of the noise around the time of a candidate event. The power spectral density is estimated from each observatory's data using Welch's method and median averaging~\cite{1161901}. The prior minute of data is divided into 4s intervals which overlap by $50\%$, and are then hann-windowed and converted to the frequency domain. The median of each frequency bin is then taken across all intervals, thus preventing short transient signals from biasing the $S_n(f)$ estimate. The median estimate is also robust to the presence of compact-binary-merger signals in the data, since the monotonic increase in frequency of such signals over time causes them to only affect one sample of any frequency bin.

The signal-to-noise time series is,
\begin{equation}
 \rho^2(t) = \frac{4}{\left\langle h | h \right\rangle} \int^\infty_0 \frac{\tilde{s}(f)\tilde{h}^*(f)}{S_n(f)} e^{2\pi ift} df
\end{equation}
which can be efficiently computed by use of the inverse fast Fourier transform. 

Fig~\ref{fig:flow} provides an overview of the filtering stages and their data use requirements. The \textit{analysis segment}, which contains the complete amount of data needed to produce a portion of uncorrupted SNR time series is shown. We can consider the action of the matched filter itself as an FIR filter with a length determined by the signal duration in the analyzed frequency band. For the analysis of O2 data, this duration is determined by the lower frequency cutoff chosen independently for each template. Using this lower frequency cutoff accumulates at least $99.5\%$ of signal power relative to using a much lower frequency cutoff\cite{DalCanton:2017ala}. A fixed number of segment sizes (typically multiples of 16 s) is chosen, allowing templates to be grouped together into sets of similar duration, and analyzed together using batched operations.

Overwhitened data is produced for each \textit{analysis segment} size by constructing an FIR filter from $1/S_n(f)$ via a truncation in the time domain and multiplying the frequency-domain data by the corresponding transfer function~\cite{Allen:2005fk}. The amount of data corrupted by this filter is discarded, introducing an additional $\sim3.5$ s of latency. We can finally compute an \textit{analysis stride} worth of valid SNR outputs for each template by way of batched inverse FFT.

\subsection{Selecting and Ranking Gravitational-wave Candidate Events}
\label{sec:rank}

In this section we will present how candidate events are identified and how they are ranked based on their signal and astrophysical consistency. We will restrict the discussion to the LIGO-Hanford and LIGO-Livingston observatories, as this was the configuration used in O2, but we note that these procedures can be extended to additional observatories. 
% discuss how we select single detector triggers 

Using the SNR time series for each template and observatory, we first identify the peak SNR in each analysis stride. This defines a single-detector \textit{trigger}. Generating triggers for \emph{each} local SNR maximum would lead to a prohibitively large number of triggers. An effective and widely-used means of controlling the number of triggers we need to analyze with minimal loss of sensitivity is to threshold on the peak SNR at $\sim5.5$. The bank of templates is divided among dozens of compute nodes for parallel analysis. Only the loudest O(10) triggers from the set of templates being analyzed by a particular compute node are kept. In all we accept O($10^3$) single detector triggers from each observatory for every analysis segment. Comparison of this procedure to the one used for the deep~\texttt{PyCBC}-based offline analysis presented in~\cite{Nitz:2017svb,Usman:2015kfa} shows this procedure is robust, and recovers $98\%$ of identified signals from a simulated population of sources.

Due to time constraints, low-latency analyses do not have access to the full set of data quality information used to determine when the detectors are in suboptimal states. In addition to making use of the minimal set of low latency data quality information, which include times of electronic control system overflows~\cite{O2-DQ-paper}, we make additional cuts based on a continuous local estimation of detector's sensitivity, the variability of the PSD over time, and the measurement of an unreasonably large SNR. These cuts serve to limit times where the instrument is clearly misbehaving in ways that would hinder the analysis, but are not currently flagged by other automated processes.

In addition, we apply a more robust signal consistency test that has been successfully employed in the deep offline analyses~\cite{Allen:2004gu,Usman:2015kfa}. For each single detector trigger we calculate a $\chi^2$ signal consistency test defined as
\begin{equation}
\chi^2_r = \frac{1}{2p-2}\sum_{i=1}^{i=p} \left\|\langle s|h_i \rangle - \langle h_i|h_i \rangle\right\|^2,
\end{equation}
where the template $h$ is divided into $p$ frequency bands which each contribute equally to the signal power. If the data, $s$, is adequately approximated by Gaussian noise with an embedded signal that is described by the waveform template $h$, this statistic will follow a reduced $\chi^2$ distribution with $2p - 2$ degrees of freedom. Many classes of non-Gaussian noise, however, tend to produce exceedingly large $\chi^2_r$ values~\cite{Allen:2004gu, Babak:2012zx}, allowing the classification of triggers as due either to astrophysical signals or non-Gaussian transient noise. This test only needs to be performed for the few SNR peaks above threshold so it is not a significant fraction of the total cost of the analysis.

We use the re-weighted SNR, introduced in~\cite{Babak:2012zx} to suppress triggers caused by non-Gaussian noise transients. This re-weighted SNR, $\tilde{\rho}$, is given as
\begin{equation}
 \tilde{\rho} = \begin{cases} 
        \rho & \mathrm{for}\ \chi^2_r <= 1 \\
        \rho\left[ \frac{1}{2} \left(1 + \left(\chi^2_r\right)^3\right)\right]^{-1/6} & 
        \mathrm{for}\ \chi^2_r > 1
    \end{cases}.
\end{equation}

To combine triggers from multiple detectors, we employ the statistic first described in \cite{Nitz:2017svb}, which includes the astrophysical probability of a trigger observed with properties $p^S(\vec{\theta})$, where $\vec{\theta}$ includes relative amplitudes, phases, and time differences between observatories. This statistic can be expressed as,
\begin{equation}\label{eq:online}
 \tilde{\rho}^2_c = \tilde{\rho}^2_H + \tilde{\rho}^2_L + 2 \ln\left(p^S(\vec{\theta})\right),
\end{equation}
where $\tilde{\rho}_L$ and  $\tilde{\rho}_H$ are the re-weighted SNRs of the triggers in the LIGO Livingston and LIGO Hanford observatories respectively. This was the statistic used by the low latency ~\texttt{PyCBC Live} analysis used in O2. In future analyses, this statistic may be extended as in \cite{Nitz:2017svb} to account for noise variation over intrinsic parameters such as the masses and spins of a binary system which was used by the ~\texttt{PyCBC}-based offline analysis, and could take into account priors over the intrinsic parameters. We further require that only single detector triggers that arise from the same template in different observatories may form a coincident candidate event. Finally, we impose a maximum 15 ms time delay between triggers from the two observatories, as the largest possible Hanford-Livingston time-of-flight is $\sim10$ ms. If there are multiple candidate events from different templates, we choose the one with the largest $\tilde{\rho}_c$ value.

\subsection{Background Estimation}
\label{sec:bkg}

\begin{figure}[h]
  \centering
    \includegraphics[width=\columnwidth]{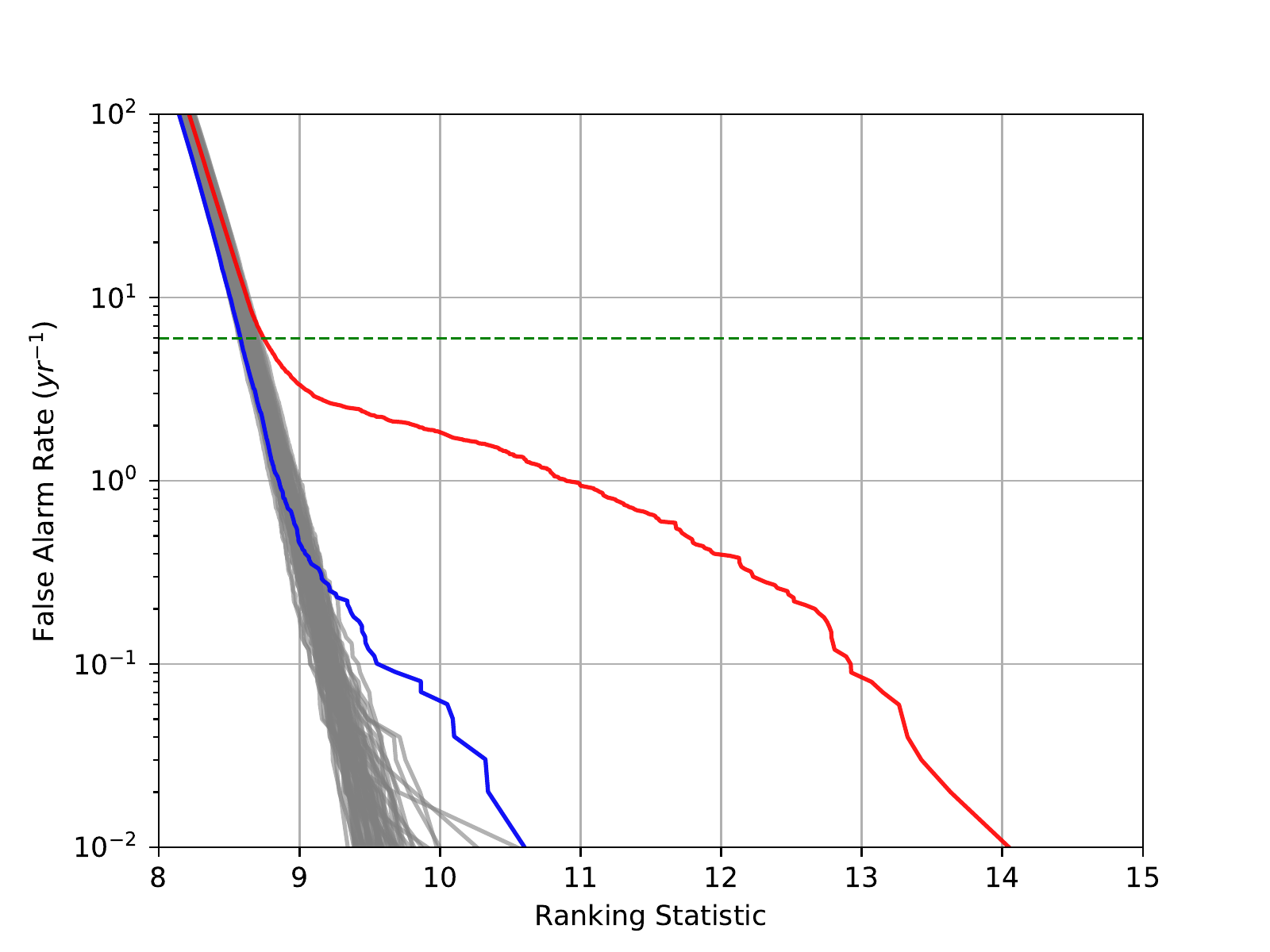}
\caption{Variability of the relationship between the ranking statistic and the false alarm rate (FAR) for independent analysis periods during O1. A green line is placed at the threshold for astronomical alerts (1 per 2 months). Background estimates were computed over 5 hour analysis periods. The mapping between the ranking statistic and the FAR is stable with the notable exception of classes of astrophysical (blue) and instrumental (red) outliers.}
\label{background_stability}
\end{figure}

As the next step, we need to determine if a candidate is significant enough for consideration by astronomers and intermediate follow-up processes. To determine the statistical significance of a candidate event, we calculate the false alarm rate of the analysis producing an event with as large a statistic value as the candidate event. The background of false coincidences is measured empirically by a resampling procedure of the data.

The data in one detector is shifted in time relative to the other detector to create a simulated analysis. Since we use time shifts that are greater than the maximum physical time-of-flight between observatories, which is 10 ms between the LIGO observatories, this analysis will only produce non-astrophysical coincidences. We repeatedly shift the data by increments of 100 ms to generate as much background as possible.  Since the analysis of each detector is independent of the time shift, we can in practice calculate the background of false coincidences by many recombinations of the single-detector triggers, which involve computationally trivial operations. This is the same procedure used by the offline analysis~\cite{Usman:2015kfa}.

For the low-latency analysis, a rolling buffer of single detector triggers is stored, which is approximately 5 hours in duration. This is used to estimate the significance of a given event down to a false alarm rate of 1 per 100 years. The limit of the inverse FAR that can be estimated, IFAR, is related to the amount of past data stored, $T_{buffer}$, and the time shift size, $T_{shift}$, by
\begin{equation}
IFAR = \frac{T_{buffer}^2}{T_{shift}}
\end{equation}
Since only 5 hours of past data is kept, the background estimate will adapt to changes in the detector state on this time scale. This is important given that the analysis will inevitably include times of highly nonstationary data, which may be later identified and vetoed from the full offline, archival analyses.

To demonstrate how different data quality affects the background, independent 5-hour data sections from the first Advanced LIGO observing run were analyzed with the~\texttt{PyCBC Live} analysis for a total of 14.8 days. A single background estimate from each was recorded and the variability of the background estimates calculated across these sections can be seen in Fig~\ref{background_stability}. Beyond a few outliers, the background estimates for independent analysis periods is stable, with the threshold for astronomical alerts (1 false alarm per 2 months) corresponding to a ranking statistic of 8.55 to 8.75.

The two cases when the background estimate is not stable are in the presence of a loud coincident event (blue) and a broadband disturbance in the PSD (red). Triggers participating in coincident events are included in the background estimate to ensure an unbiased rate of false alarms~\cite{Capano:2016uif}. Similar behavior is seen in estimates from the high-latency searches~\cite{Usman:2015kfa,Capano:2016uif}.

Due to the low-latency nature of this pipeline, fewer data quality products are available to be utilized by the search. The data quality information distributed in low-latency includes only short term disturbances that are often vetoed by the consistency checks the search incorporates~\cite{TheLIGOScientific:2017lwt}. Data quality products that indicate broadband biases to the PSD, which generally necessitate the removal of that time from analyses, are not processed in low-latency, and hence not available to this search. These disturbances lead to an increase in the rate of loud non-coincident events that are incorporated into the background. In O2, the amount of time that the low-latency search processed, but was not searched by the ~\texttt{PyCBC}-based offline analysis due to additional data quality inputs is approximately 2.1 days, indicating that ~$2\%$ of data which the low-latency search would analyze is later thrown out in the offline analysis~\cite{O2-DQ-paper}.
%This difference is based on the coincident time from only using the vetoes that pycbc live takes advantage of (all else being equal)
% 119.07 days vs 116.95 days

\subsection{Architecture and Computational Considerations}

\begin{figure}[h]
  \centering
    \includegraphics[width=\columnwidth]{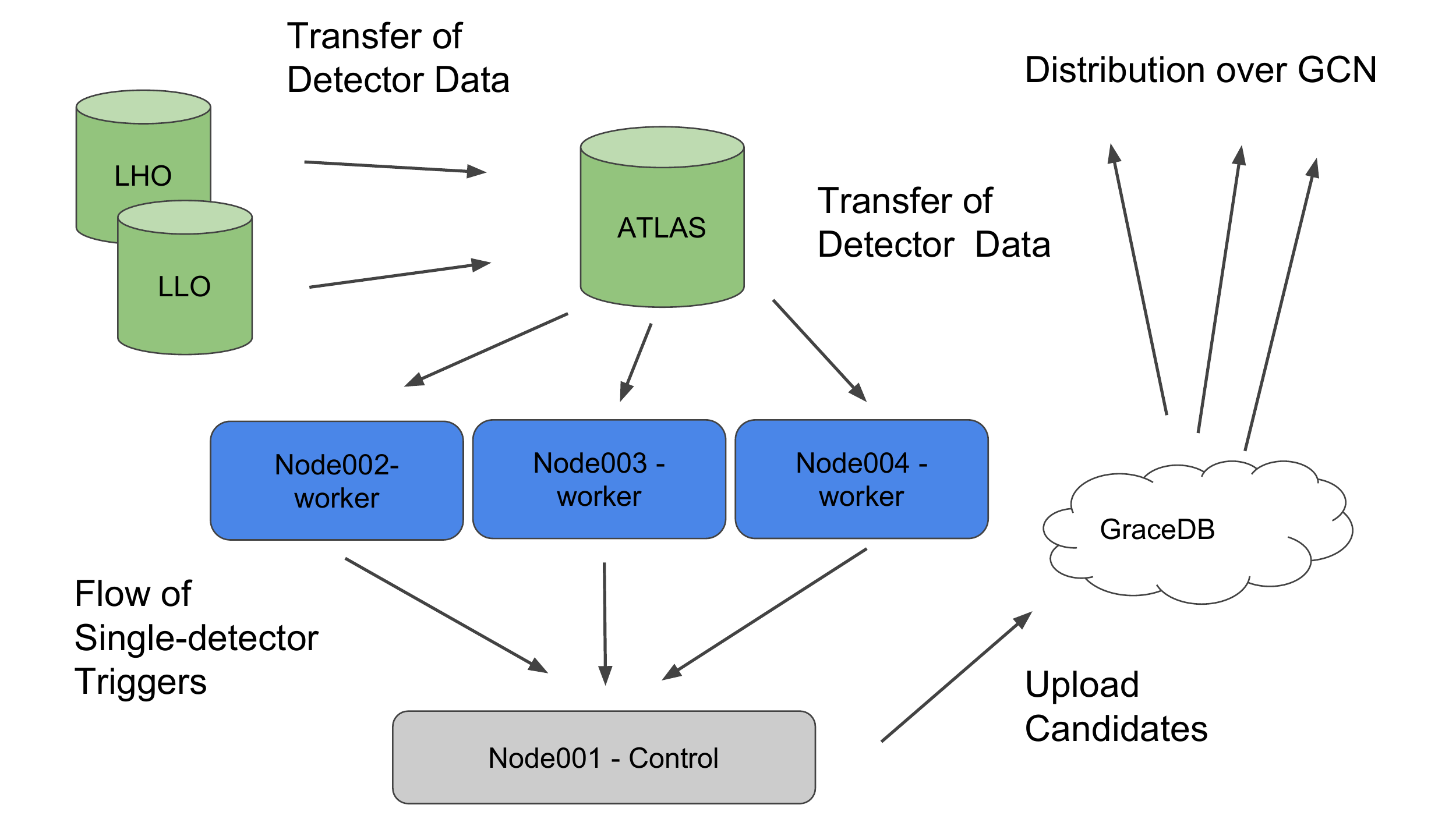}
\caption{The high level overview of how data is flowing from the observatories
through the \texttt{PyCBC Live} analysis, to GraceDB\cite{GraceDB}, and finally to astronomers by way of the Gamma-ray Coordinates Network (GCN). The work nodes (blue) process different portions of a template bank, while the control node (gray) collates all these results and determines if there is a significant candidate event.}
\label{fig:arch}
\end{figure}

In this section we will discuss the high level architecture of the \texttt{PyCBC Live} analysis, along with the computational cost and scaling. Fig~\ref{fig:arch} provides a high level diagram of how the analysis is ordered. We see that data from the observatories is first distributed to computing clusters, such as the ATLAS supercomputer~\cite{atlas}, onto cluster nodes via multicast.

Once data is distributed to the cluster nodes, the \texttt{PyCBC Live} analysis takes over and calculates the SNR time series for each template and observatory, which is a highly parallel process. To ensure the analysis completes in a time shorter than the \textit{analysis stride}, we distribute the work over multiple computing nodes by use of task parallelization via Message Passing Interface (\texttt{MPI}~\cite{mpi4py}), represented in blue in Fig~\ref{fig:arch}.

In the early advanced-detector era, template banks typically contain O($10^5$) templates. However, template waveforms are sufficiently short that they can be generated once at the beginning of an analysis and stored in system memory indefinitely. We see in Fig~\ref{fig:flow} how the duration of the template affects the length of an individual \textit{analysis segment}. Templates are grouped in batches of similar duration and analyzed together. This allows for the calculation of the SNR time series to be further parallelized on each compute node using \texttt{OpenMP} by taking advantage of the well optimized batched FFT algorithms provided by \texttt{FFTW}~\cite{Frigo05thedesign} and Intel's \texttt{MKL} library.

Conceptually, each compute node handles its own section of the template bank independently of the others and produces its own set of single detector triggers. This computation comprises the vast majority of the overall computing cost of the analysis. It is efficient to simply transfer the recorded triggers from all the nodes to a single control node (gray in Fig~\ref{fig:arch}) where triggers from separate
observatories are combined into candidate events, ranked, and assigned a statistical
significance after the background is estimated. This same process is responsible for
finally uploading a candidate event to the Gravitational-wave Candidate Database (\texttt{GraceDB})~\cite{GraceDB} if it passes nominal false alarm rate thresholds (typically 1 / day). 

To ensure that the analysis keeps up with the incoming data, it is sufficient to simply use enough worker nodes such that it completes the previous set of analysis segments before it must begin the next. In practice, the analysis is configured to only take $80\%$ of this time. During an observing run there are disruptions in the data stream, or other delays, which cause the data distribution and analysis to fall behind. When the disruption subsides, this extra headroom allows the analysis to catch up rapidly.

A common measurement of computational performance for a template-based gravitational-wave search is how many templates a CPU core can analyze in real time. During O2 this analysis achieved an in-situ performance of 5000 templates per core at $80\%$ load, or ~6300 templates per core at full machine load. The computational scaling of this analysis is related to the length of templates which need to be analyzed, and the latency desired. For a similar analysis to the one performed in O2, reducing the \textit{analysis stride} from 8 to 4 seconds, decreases the average analysis-induced latency from 16 to 10 seconds, while increasing the computational cost by just under a factor of 2. This occurs because the computational cost $C$ has roughly the following scaling relation,
\begin{equation}
C \propto \frac{\bar{T}\ln(\bar{T})}{T_{stride}}
\end{equation}
where $\bar{T}$ is the template duration averaged over the template bank, and $T_{stride}$
is the \textit{analysis stride}. $\bar{T}$ is primarily determined by the shape of the PSD during a particular observing run. There is a balance between the latency of the analysis and the computational cost. Also note that a reduction in the analysis stride below a certain value no longer results in a proportionate drop in the average overall latency, due to unavoidable latency incurred from data transfers and conditioning filters.

\section{Sensitivity of the Analysis}
\label{sec:vt}

\begin{figure}[h]
\centering
    \includegraphics[width=\columnwidth]{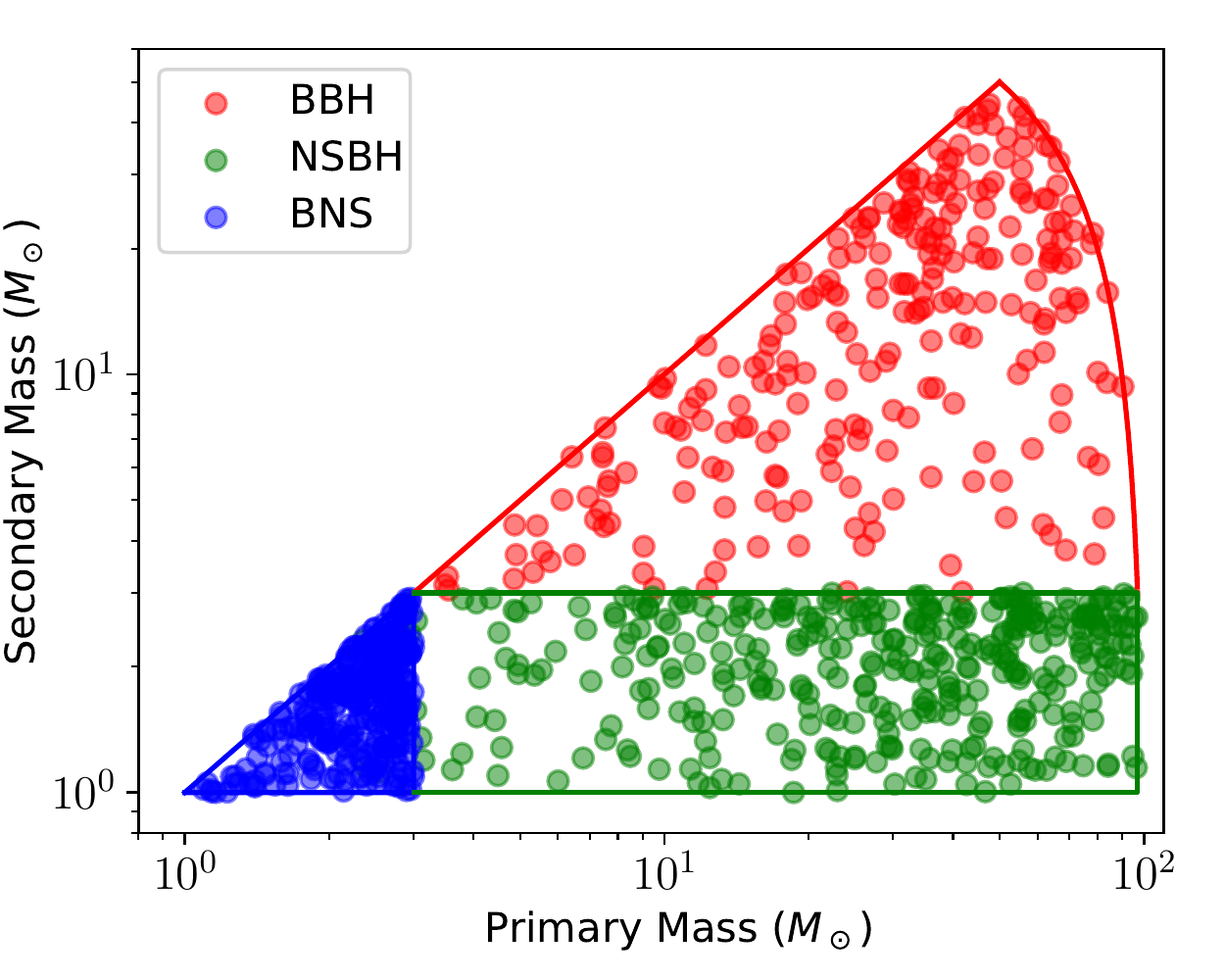}
\caption{Binary component masses of the samples of the offline analysis injection set. BNS population is constrained between component masses $m_1$, $m_2$ $\epsilon$ (1.0, 3.0) $M_\odot$, the NSBH population ranges between $m_1$ $\epsilon$ (1.0, 3.0), $m_2$ $\epsilon$ (1.0, 97.), and the BBH population ranges between $m_1$, $m_2$ $\epsilon$ (1.0, 99.0) and $m_2 + m_1 < 100$.}
\label{selected_injs}
\end{figure}

The sensitivity of a gravitational-wave search is estimated by simulating an astrophysical population of sources, adding the signals to a data set, and observing how many signals are detected. The source parameters of each simulated signal are selected from three binary populations that are expected to be the most significant sources of gravitational-waves from compact binary mergers: binary neutron stars (BNS), binary black holes (BBH), and neutron star-black hole binaries (NSBH). We model these sources using an effective-one-body, inspiral-merger-ringdown waveform model for NSBH and BBH signals and a post-Newtonian, inspiral-only model for BNS systems~\cite{Blanchet:2002av,Faye:2012we,Taracchini:2013rva,Bohe:2016gbl}. Once these signals are added into detector data from LIGO's first observing run, the full~\texttt{PyCBC Live} search is conducted and signals are recovered with an estimated false alarm rate. The sensitivity of the analysis can be determined by a weighted Monte Carlo integration of the distances at which sources are found above a given detection threshold~\cite{TheLIGOScientific:2016qqj, Capano:2016dsf}. 

Here we take a subset of the software injections used to evaluate the offline \texttt{PyCBC} analysis~\cite{Nitz:2017svb} and compare the sensitivity of the low latency ~\texttt{PyCBC Live} and ~\texttt{PyCBC}-based offline analyses. To minimize differences where possible, we consider the same detection statistic for both, however, we note that the offline analysis
used in the first observing run of Advanced LIGO treated background separately for templates corresponding to mergers with total mass less than four solar masses~\cite{TheLIGOScientific:2016qqj}, whereas the \texttt{PyCBC Live} analysis treats them together. We expect this difference to increase the sensitivity of the offline analysis to lower-mass sources. We consider O(1000) BNS, BBH, and NSBH-like sources. These injections sufficiently cover the parameter space investigated during the O1 offline analysis as can be seen in Figure \ref{selected_injs}. The source orientation and sky location is chosen isotropically, and sources are weighted based on their distance so as to achieve an effective uniform-in-volume distribution. We choose to evaluate the sensitivity at a false alarm rate of 1 per 2 months, to match the threshold used for alert generation by LIGO/Virgo. Table~\ref{table:vt} shows that despite the differences in the \texttt{PyCBC Live} and \texttt{PyCBC}-based offline analysis the sensitivity is comparable and matches our expectation.

\begin{table}
\begin{tabular}{ |c|c|c| } 
\hline
Source Category & Relative Sensitivity\\
\hline
BNS & $0.87 \pm 0.05$\\ 
NSBH & $0.98 \pm 0.04$\\ 
BBH &   $1.05 \pm 0.09$\\
\hline
\end{tabular}
\caption{Relative sensitivity of the \texttt{PyCBC Live} low-latency analysis and the \texttt{PyCBC}-based offline analysis for different source classes evaluated using data from LIGO's first observing run, and using a false alarm rate threshold of 1 per 2 months. We see that the \texttt{PyCBC Live} analysis is close to the sensitivity of the full offline analysis which benefits from improved data quality information, improved calibration, and a larger set of signal consistency tests~\cite{Nitz:2017svb,Nitz:2017lco}.}
\label{table:vt}
\end{table}

\section{Multidetector sky localization}

The addition of a third instrument to the detector network dramatically reduces the uncertainty on the sky localization of candidate events, in some cases going from hundreds to tens of square degrees \cite{Singer:2014qca}. The additional detector does not need to be as sensitive as the others to provide a noticeable contribution, and in fact does not need to detect the candidate at all, as demonstrated by GW170817. The rapid advance of the Virgo commissioning during 2017, therefore, prompted the need to include its data in the rapid localization of low-latency candidates from \texttt{PyCBC Live}.

An optimal search of three-detector data involves matched-filtering all detectors separately and combining the result in a way that varies over a grid of many possible source sky locations \cite{Macleod:2015jsa}. Thus, its implementation is considerably more complex than the two-detector coincidence algorithm described earlier. However, if two of the detectors have a comparable sensitivity and the third is much less sensitive---a configuration expected for the LIGO/Virgo network towards the end of O2---applying the simple coincidence to the first two detectors and ignoring the third provides a reasonable approximation to the optimal search in terms of detection rate \cite{Macleod:2015jsa}.

The rapid localization of candidates identified in low latency is performed by the \texttt{BAYESTAR} algorithm \cite{Singer:2015ema}. The most recent version of \texttt{BAYESTAR} takes as input a candidate's complex SNR time series and a local estimate of the noise PSD from any number of detectors, regardless of which detectors actually reported a trigger for that candidate. Therefore, \texttt{PyCBC Live} proceeds as follows. If a candidate from the LIGO detectors is found significant enough to submit to \texttt{GraceDB}, we check the availability and quality status information of Virgo data for the entire duration of the template that produced the candidate. If the checks pass, we calculate the Virgo complex SNR time series around the time of the candidate, using the same template. This incurs negligible computation cost, since it involves a single template and less than 0.1s of data. The resulting set of SNR time series from LIGO and possibly Virgo is handed over to \texttt{BAYESTAR} to generate the low-latency localization, regardless of the SNR of the candidate in Virgo. The significance of the candidate does not use the Virgo SNR and is based on LIGO data only. In this way, we can rapidly produce a three-detector localization without having to calculate a three-detector ranking statistic and estimate its background distribution. Note that the same procedure is also applied to significant single-detector LIGO candidates (see Sec.~\ref{sec:single}), such that their localization is actually based on two detectors, if Virgo data are available.

Virgo is expected to be significantly more sensitive in O3 and will thus require a three-detector ranking statistic in order for the search to achieve its maximum sensitivity. Nevertheless, the technique described above can still be applied for new detectors---such as KAGRA---which may join the network with an initially reduced sensitivity as their commissioning proceeds.

\begin{figure}
    \includegraphics[width=\columnwidth]{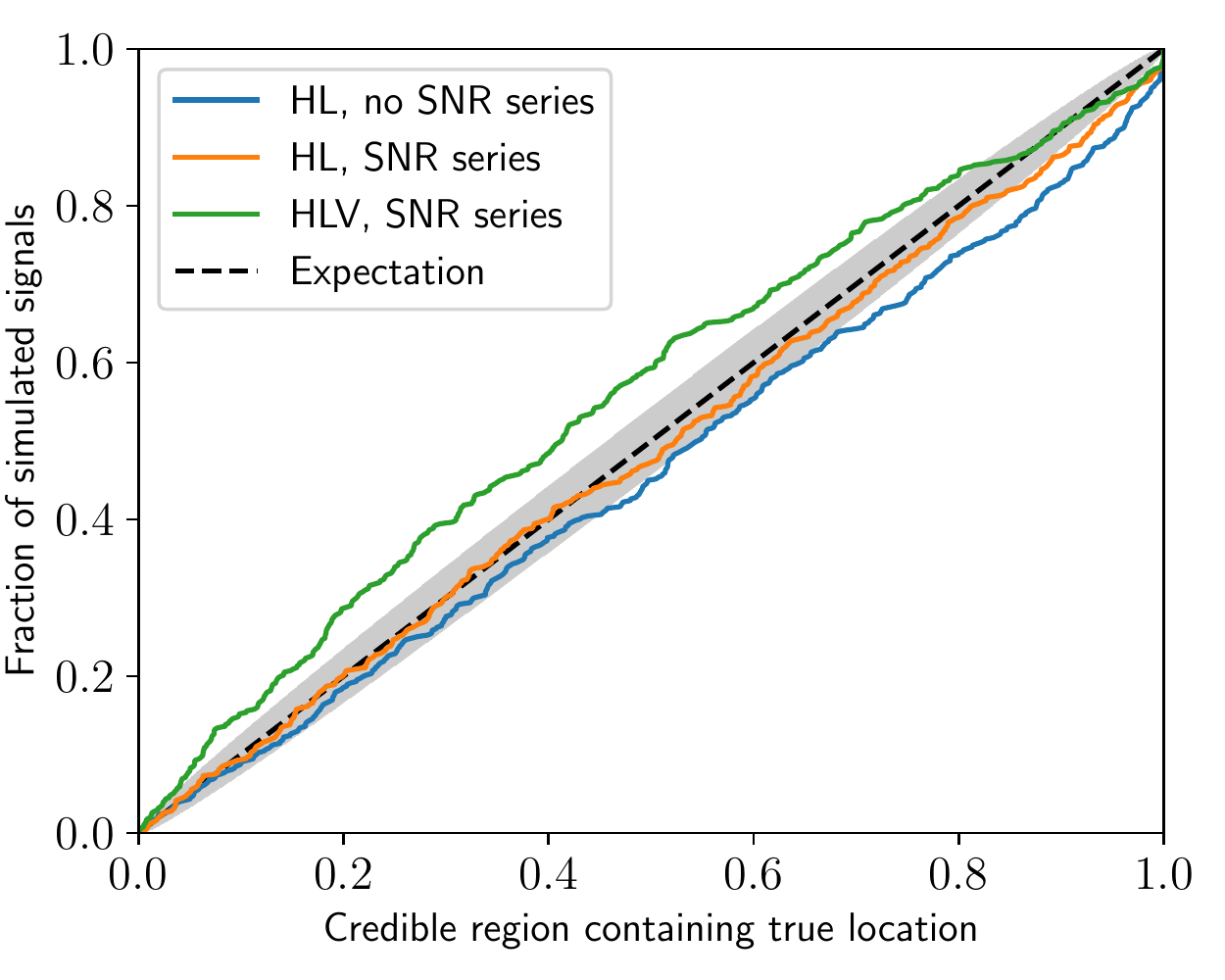}
    \caption{Reliability of the rapid sky localization of simulated signals detected by \texttt{PyCBC Live} and localized by \texttt{BAYESTAR}. The blue curve shows signals detected and localized using the LIGO Hanford (H) and Livingston (L) detectors only, and only using point estimates of the arrival time, arrival phase and SNR at the two detectors. In the orange curve, the point estimates are replaced by the full SNR time series from the two detectors. In the green curve, the SNR time series from Virgo (V) is included as well, regardless of the presence of a Virgo trigger. The gray band is the binomial 95\% interval given the number of detected simulations. The fraction of signals whose true location falls within a given credible region is close to the expectation, although the HLV localizations tend to be slightly larger than expected.}
    \label{fig_bayestar_pp}
\end{figure}

We test the described multidetector localization procedure by simulating 3000 BNS, NSBH and BBH signals immersed in Gaussian noise, detecting them with \texttt{PyCBC Live} and generating localization maps of the resulting candidates as described. Waveform models, similar to those used in Sec.~\ref{sec:vt} are used, and all sources are distributed uniformly in volume, sky location and spatial orientation. The detector noise is simulated from analytic PSD models, qualitatively comparable to estimates from  the tenth engineering run of the LIGO detectors, and a factor of $\sim 3$ less sensitive for Virgo\footnote{Specifically the aLIGOMidLowSensitivityP1200087 and AdVEarlyLowSensitivityP1200087 models from LALSimulation \cite{lalsuite}.}. Such models are chosen to represent the sensitivity of the detectors at the end of O2. The reliability of the produced localization maps can be tested by plotting the fraction of simulated signals whose true location falls within a certain probability of the posterior sky location. Figure~\ref{fig_bayestar_pp} shows such a plot for three configurations: (i) the \texttt{BAYESTAR} algorithm employed during O1 and early O2, which only used point estimates of the candidate event time, amplitude and phase instead of the full SNR time series (blue), (ii) \texttt{BAYESTAR} in its late-O2 version, using the full SNR time series, but from LIGO detectors only (orange), and (iii) \texttt{BAYESTAR} using the full LIGO and Virgo SNR time series (green).

We find that using the SNR time series leads to consistent sky localizations, although the posteriors with Virgo are slightly conservative: about 60\% of the 50\% credible regions contain the simulated location. We attribute this small discrepancy to a tunable parameter in the \texttt{BAYESTAR} algorithm, which we have left at the default value for the purpose of this test and could in principle be optimized to achieve exactly consistent posteriors for the three-detector network as well. We conclude that the standard \texttt{BAYESTAR} settings produce consistent localizations for events observed by LIGO only, and slightly conservative localizations for events observed by Virgo as well. We find a typical reduction in uncertainty with the inclusion of Virgo to be roughly an order of magnitude, similarly to that found in \cite{Singer:2014qca}.

\section{Other Applications}

\subsection{Single Detector Search}
\label{sec:single}

We have focused on the detection of gravitational-waves that are observed in multiple detectors. It may be difficult to claim a detection based solely on a single detector analysis, due to the background incurred from a detector's non-Gaussian noise transients. However, as the sensitivity of the instrument increases, and the detection rate increases, it has been found that some of the most significant single detector triggers may be astrophysical in origin~\cite{Callister:2017urp}. Maximizing the scientific output of a compact binary merger requires electromagnetic observations as close to the merger as possible~\cite{ArcaviKilonova}, so rapidly identifying and distributing the most promising candidates when only a single detector is operating is warranted. Unfortunately, sky localization is a significant challenge in this case from gravitational-wave observation alone. On the other hand, observations of associated prompt gamma-ray emissions, for instance with wide-field observatories like Fermi/GBM~\cite{FermiGBM}, may be able to localize the event with sufficient precision for subsequent followup.

% Citations for above???

To allow for the detection of gravitational waves from compact binary mergers even when only a single detector is running, we can employ most of the tools we have already described. The ranking of gravitational-wave triggers at a single detector level, proceeds in the same manner. However, a few additional cuts are made to help cope with the decreased ability to reject noise. Candidates which are consistent with very short-duration signals are excluded, due to the difficulty in separating short duration transient noise from high mass binary black hole mergers~\cite{Nitz:2017lco}. In addition, our signal consistency test is more stringently applied and a cut directly on the reduced $\chi^2<4$ is used. The primary goal is to enable the detection of binary neutron star mergers, which are well modelled by the templates and chirp for a relatively long time, making our signal consistency tests particularly effective. In addition, binary neutron star candidates have the highest chance of being aided in sky localization by coincident wide-field gamma-ray observations. Finally, candidates are chosen and distributed based on an empirically-measured threshold which produces false alarms at a rate of less than 1 / month.

\subsection{Data Monitoring}

The \texttt{PyCBC Live} analysis was employed and ran continuously through the second observing run of Advanced LIGO, including the period of coincident observation with Virgo. Along with generating detection alerts, it was used as a low-latency monitoring tool of the data. Triggers, which include information such as time, signal-to-noise, template masses and spins, along with signal consistency tests, are displayed through the LIGO detector characterization summary pages~\cite{gwsumm}. These are regularly monitored by analysts for anomalous behavior, which could be a sign of adverse detector behavior. This regular monitoring has also been critical to the detection of two gravitational-wave events during O2; both GW170104 and GW170608 were initially discovered by visual inspection of \texttt{PyCBC Live} results from these monitoring pages~\cite{GW170608GCN,GW170104GCN}. In particular, GW170608, was discovered during time where the LIGO-Hanford instrument was in a commissioning phase~\cite{Abbott:2017gyy}.

\section{Conclusions}

In this paper we have presented \texttt{PyCBC Live}, a simple and effective analysis designed to rapidly detect gravitational-waves from compact binary mergers, built upon the existing technologies used for the well-established offline analysis~\cite{Usman:2015kfa}, and distributed as free and open software in the~\texttt{PyCBC} library~\cite{pycbc-github}. \texttt{PyCBC Live}'s successful operation during O2, where it has been involved in either the detection or analysis of all currently published gravitational-wave events~\cite{GW170814GCN, GW170104GCN, GW170608GCN, GCN21513}, demonstrates that well-known, standard frequency-domain matched filtering is a successful technique for the low-latency detection of gravitational waves. 

We have primarily focused on the coincident detection of a gravitational-wave observations at the two current LIGO interferometers. In the coming years, the Virgo detector will improve in sensitivity, while KAGRA and LIGO India will also join the worldwide network of gravitational-wave observatories~\cite{Aasi:2013wya}. The analysis can already produce accurate sky localizations from any number of additional observatories. The analysis can also be straightforwardly extended to analyze more than two detectors in a symmetric manner. A similar technique to the multidetector compact binary merger analysis performed in the fifth and sixth observing runs of initial LIGO/Virgo~\cite{Babak:2012zx} will be investigated in the future.

\acknowledgments
We thank Thomas Dent, Duncan Brown, and Collin Capano for their input, Leo Singer for his help with \texttt{BAYESTAR}, along with Laura Nuttall, Gregory Mendell and Alan Weinstein for their useful comments and feedback during the internal LIGO Scientific Collaboration review process. 

We acknowledge the Max Planck Gesellschaft for support and the Atlas cluster computing team at AEI Hannover. Computations were also supported by Syracuse University and NSF award OAC-1541396. We also thank the LIGO Scientific Collaboration for access to the data and gratefully acknowledge the support of the United States National Science Foundation (NSF) for the construction and operation of the LIGO Laboratory and Advanced LIGO as well as the Science and Technology  Facilities Council (STFC) of the United Kingdom, and the Max-Planck-Society (MPS) for support of the construction of Advanced LIGO. Additional support 
for Advanced LIGO was provided by the Australian Research Council.

TDC was supported by an appointment to the NASA Postdoctoral Program at the
Goddard Space Flight Center, administered by Universities Space Research
Association under contract with NASA.

\bibliography{references}

\end{document}